\documentclass[aps,prc,amsfonts,preprintnumbers,superscriptaddress,showpacs,nofootinbib]{revtex4}

\usepackage{graphics}
\usepackage{psfrag}
\usepackage{epsfig}
\usepackage{epsf}
\usepackage{float}

\newcommand{\fet}[1]{\mbox{\boldmath $#1$}}

\newcommand{\beq}{\begin{equation}}
\newcommand{\eeq}{\end{equation}}
\newcommand{\beqa}{\begin{eqnarray}}
\newcommand{\eeqa}{\end{eqnarray}}
\newcommand{\nn}{\nonumber \\ }

\def\palka{\hspace{-7.0pt}/ \,}

\begin{document}

\preprint{FZJ-IKP-TH-2007-9}
\preprint{HISKP-TH-07/06}

\title{Nuclear forces with $\Delta$-excitations up to next-to-next-to-leading 
order I:\\[0.2em]
peripheral nucleon-nucleon waves}

\author{H.~Krebs}
\email[]{Email: hkrebs@itkp.uni-bonn.de}
\author{E.~Epelbaum}
\email[]{Email: e.epelbaum@fz-juelich.de}
\affiliation{Forschungszentrum J\"ulich, Institut f\"ur Kernphysik 
(Theorie), D-52425 J\"ulich, Germany}
\affiliation{Universit\"at Bonn, Helmholtz-Institut f{\"u}r
  Strahlen- und Kernphysik (Theorie), D-53115 Bonn, Germany}
\author{Ulf-G.~Mei{\ss}ner$\,^{2,1}$}
\email[]{Email: meissner@itkp.uni-bonn.de}
\homepage[]{URL: www.itkp.uni-bonn.de/~meissner/}
\date{\today}

\begin{abstract}
We study the two-nucleon force at next-to-next-to-leading order in a chiral
effective field theory with explicit $\Delta$ degrees of freedom. Fixing the
appearing low-energy constants from a next-to-leading order calculation of
pion-nucleon threshold parameters, we find an improved convergence of most
peripheral nucleon-nucleon phases compared to the theory with pions and nucleons
only. In the delta-full theory, the next-to-leading order corrections are 
dominant in most partial waves considered. 
\end{abstract}


\maketitle

\vspace{-0.2cm}

\section{Introduction}
\def\theequation{\arabic{section}.\arabic{equation}}
\label{sec:intro}

The forces between nucleons based on chiral effective field theory have been 
studied in great detail over the last decade, for reviews see e.g.
\cite{Bedaque:2002mn,Epelbaum:2005pn}. Most calculations are based on
the chiral effective Lagrangian formulated in terms of the asymptotically
observed ground state fields, the pions and nucleons chirally coupled 
to external sources. The excitation of baryon and meson resonances is encoded 
in the low-energy constants (LECs) of the pion-nucleon interaction beyond 
leading order. Such a framework provides an accurate representation of the 
nucleon-nucleon (NN) phase shifts if extended to sufficient high order,
state-of-the-art calculations have been carried out to
next-to-next-to-next-to-leading order in the chiral expansion. This is
very similar to the description of pion-nucleon scattering in chiral
perturbation theory well below the $\Delta$-excitation energy, where a
precise description of the pertinent phase shifts has been obtained in 
a complete one-loop (fourth order) calculation, see e.g. Ref.~\cite{Fettes:2000xg}.
Still, it can be argued that the explicit inclusion of the delta, which 
is the most important resonance in nuclear physics, allows
one to resum a certain class of important contributions and thus leads
to an improved convergence as compared to the delta-less theory, provided
a proper power counting scheme such as the small scale expansion (SSE) 
\cite{Hemmert:1997ye} is employed. The SSE is a phenomenological extension
of chiral perturbation theory in which the delta-nucleon mass splitting is
counted as an additional small parameter. This improved convergence has been
explicitely demonstrated for pion-nucleon scattering where the description of the 
phase shifts at third order in the SSE comes out superior (inferior) to the
third (fourth) order chiral expansion in the pure pion-nucleon theory
 \cite{Fettes:2000bb}. Note, however, that the theory with explicit deltas
has more LECs at a given order due to the richer operator structure when 
spin-3/2 fields are present. Clearly, for a description of pion-nucleon scattering 
in the $\Delta$-region or pion production in NN collisions the 
inclusion of the spin-3/2 fields as active degrees of freedom is mandatory,
see e.g.  \cite{Hanhart:2003pg} for a review.

In this work, we want to analyze the two-nucleon forces by systematically 
including  the  $\Delta$-resonance beyond leading order in the
small scale expansion, extending earlier work presented in 
Refs.~\cite{Ordonez:1995rz,Kaiser:1998wa} (the precise relation to these
papers will be given below). More precisely, we construct and analyze the
two-pion exchange NN potential at next-to-next-to-leading order (NNLO). 
For that, we must determine the sum of the subleading
$\pi N \Delta$ LECs $b_3, b_8$ together with the $\pi N$ LECs $c_i$ (the
values of these differ from the ones obtained in chiral perturbation theory).
This is achieved by working out the threshold coefficients of pion-nucleon 
scattering from tree graphs at second order 
which is of sufficient accuracy for the inclusion of the corresponding
operators in the NNLO potential. From that, we can deduce the momentum and coordinate
space representations of various isoscalar and isovector potentials.
Having done that, we calculate the peripheral
phases in NN scattering based on perturbation theory. It is well established
that these peripheral waves do not require a non-perturbative resummation and
therefore let one most directly analyze the effects of various contributions
to the effective two-nucleon potential. 

The manuscript is organized as follows. In Sect.~\ref{sec:2NF} we work out the
contributions of the two-pion exchange (TPE) graphs including the $\Delta$-resonance
at NNLO. We then determine the corresponding dimension-two LECs from a fit
to the $\pi N$ threshold parameters in Sect.~\ref{sec:LEC}. The resulting 
coordinate space representation of the TPE potential and the peripheral
NN waves are shown and discussed in Sect.~\ref{sec:results}. We end with a
summary and outlook. The appendix contains explicit analytical formulae for
the  $\pi N$ threshold coefficients.


\section{$\Delta$-contributions to the two-nucleon force up to NNLO}
\def\theequation{\arabic{section}.\arabic{equation}}
\label{sec:2NF}

In this chapter, we construct the two-pion exchange potential (TPEP)
with one or two intermediate $\Delta$-state(s) at NNLO. 
We employ here standard Weinberg power counting for the two-nucleon
effective potential, that is the leading TPEP starts at 
next-to-leading order (NLO) ($\nu = 2$)
and the first corrections to it appear at NNLO ($\nu = 3$). The leading
order ($\nu = 0$) potential is given by the static one-pion exchange. Since we
are only considering peripheral waves with the angular momentum $l \geq 2$, 
no four-nucleon contact interactions contribute to the accuracy we are working at. 

\subsection{Effective Lagrangian and power counting}
The calculations performed in the following are based on the effective
chiral Lagrangian of pions, nucleons and deltas. We employ here the
heavy baryon formulation and display only the terms of relevance for
our study:
\beqa
{\cal L} &=& {\cal L}^{(1)} + {\cal L}^{(2)} + \ldots  \nonumber\\
{\cal L}^{(1)} & = & \bar{N} \, {\cal A}_{\pi N}^{(1)} \, N
                 +   \bar{T} \, {\cal A}_{\pi \Delta}^{(1)} \, T  
                 +   \left(\bar{T} \, {\cal A}_{\pi N \Delta}^{(1)} \, N 
                     + {\rm h.c.}\right)~,
\nonumber\\
{\cal L}^{(2)} & = &  \bar{N} \, {\cal A}_{\pi N}^{(2)} \, N
    + \left( \bar{T} \, {\cal A}_{\pi N \Delta}^{(2)} \, N + {\rm h.c.}\right)~,
\eeqa
with
\beqa
{\cal A}_{\pi N}^{(1)} &=& i \, v\cdot D + g_A \, u \cdot S~,  \nonumber\\
{\cal A}_{\pi N}^{(2)} &=& c_1 \, \langle \chi_+ \rangle + c_2 \, (v \cdot
       u)^2 + c_3 \, u \cdot u + c_4 \, [ S^\mu, S^\nu ] u_\mu u_\nu ~,
                                                               \nonumber\\
{\cal A}_{\pi \Delta}^{(1)} &=& - \, \left( i v\cdot D^{ij} - (m_\Delta -
      m_N) \, \delta^{ij} + \ldots \right)\, g^{\mu \nu} ~,  \nonumber\\
{\cal A}_{\pi N\Delta}^{(1)} &=& h_A \, P^{\mu\alpha} \, w_\alpha^i~, \nonumber\\
{\cal A}_{\pi N\Delta}^{(2)} &=& P^{\mu\alpha} \, i \, (b_3+b_8) \, 
              w_{\alpha\beta}^i \, v^\beta~,    
\eeqa
where $N$ denotes the large component of the nucleon field, $T$ is an
abbreviation for the large component of the delta field, $T \equiv T_\mu^i$,
with $i$ an isospin and $\mu$ a Lorentz index. We use standard notation:
$U(x) = u^2(x)$ collects the pion fields, $u_\mu = i(u^\dagger \partial_\mu u
- u \partial_\mu u^\dagger)$, $\chi_+ = u^\dagger \chi u^\dagger + u
\chi^\dagger u$ includes the explicit chiral symmetry breaking,
$\langle \ldots \rangle$ denotes a trace in flavor space and $D_\mu
(D_\mu^{ij})$ is the chiral covariant derivative for the nucleon (delta) 
fields. Furthermore,
$P_{\mu \nu}$ is the standard projector on the 3/2-components, 
$P_{\mu \nu} = g_{\mu \nu} - v_\mu v_\nu - 4S_\mu S_\nu /(1-d)$, 
with $v_\mu$ the four-velocity, $S_\mu$ the covariant spin vector and 
$d$ the number of space-time dimensions. We also have $w_\alpha^i = 
\langle \tau^i u_\alpha \rangle /2$ and  $w_{\alpha\beta}^i =
\langle \tau^i [\partial_\alpha , u_\beta] \rangle / 2$. 
In the following, we also use the notation $\Delta \equiv m_\Delta - m_N$ for
the $N\Delta$ mass splitting (which can not be confused with the same symbol
denoting the delta field).
For further notation and discussion, we refer
to Ref.~\cite{Fettes:2000bb}.  The pertinent LECs are at leading order
the nucleon axial-vector coupling $g_A$ and the $\pi N \Delta$ axial coupling
$h_A$. At NLO, we have the four dimension-two $\pi N$ LECs $c_i$ ($i=1,2,3,4$)
and the combination of $\pi N\Delta $ LECs $b_3 +b_8$. Strictly speaking,
all couplings and masses appearing in the effective Lagrangian should be
taken at their chiral limit values, but to the accuracy we are working, 
we can use their pertinent physical values.

The various terms are ordered according to the so-called small scale
expansion, in which the small expansion parameter $Q$ includes external momenta, 
pion masses and the nucleon-delta mass splitting,
\beq
Q \in \{ p/\Lambda_\chi, M_\pi/ \Lambda_\chi, (m_\Delta - m_N)/\Lambda_\chi \}~,
\eeq
with $\Lambda_\chi \simeq 1\,$GeV the scale of chiral symmetry breaking. 
The various contributions are ordered according to the chiral power $\nu \geq
0$. Note
also that in the Weinberg power counting used here, $1/m$ corrections to 
vertices and propagators are suppressed by an extra power of $Q$. We remark that
while the SSE is a phenomenologically viable extension of chiral perturbation
theory, it can not be straightforwardly used to study the chiral limit of QCD. For
that, one has to enforce decoupling. We refrain from a more detailed
discussion of this issue here since it does not play a role for the results
presented in this paper.

\subsection{Two-pion exchange potential with intermediate deltas}

\begin{figure}[tb]
  \begin{center} 
\includegraphics[width=13.0cm,keepaspectratio,angle=0,clip]{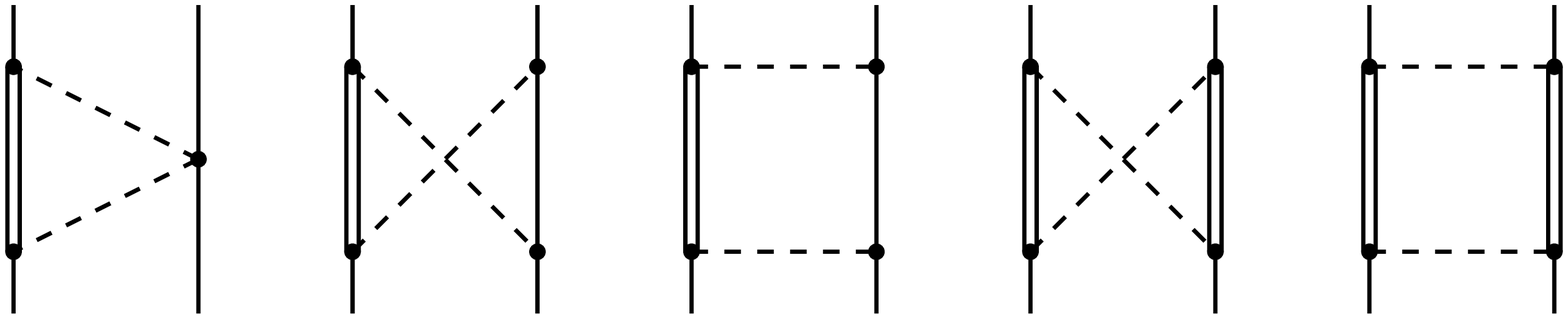}
    \caption{Leading contributions to the $2\pi$--exchange NN potential with single and 
double $\Delta$--excitations. Dashed, solid and double-solid lines represent pions,
nucleons and $\Delta$--isobars, respectively. Solid dots denote the 
leading order (LO) vertices.  
    \label{nlo} 
 }
  \end{center}
\end{figure}

The NN potential in
the center-of-mass system (CMS) can be conveniently expressed in the form:
\beqa
\label{2PEdec}
V &=& V_C + \fet \tau_1 \cdot \fet \tau_2 \, W_C + \left[   
V_S + \fet \tau_1 \cdot \fet \tau_2 \, W_S \right] \, \vec \sigma_1 \cdot \vec \sigma_2 
+ \left[ V_T + \fet \tau_1 \cdot \fet \tau_2 \, W_T \right] 
\, \vec \sigma_1 \cdot \vec q \,\, \vec \sigma_2 \cdot \vec q \\
&+&    \left[   
V_{LS} + \fet \tau_1 \cdot \fet \tau_2 \, W_{LS} \right] \, i ( \vec \sigma_1 + \vec \sigma_2 )
\cdot ( \vec q \times \vec k  ) 
+   \left[   
V_{\sigma L} + \fet \tau_1 \cdot \fet \tau_2 \, W_{\sigma L} \right] \,   \vec \sigma_1 
\cdot (\vec  q \times \vec k  ) \,\, \vec \sigma_2 \cdot (\vec  q \times \vec k  ) \,,
\nonumber
\eeqa
where the superscripts $C$, $S$, $T$, $LS$ and $\sigma L$ of the scalar functions 
$V_C$, $\ldots$, $W_{\sigma L}$ refer to the central, spin-spin, tensor, spin-orbit and 
quadratic spin-orbit components, respectively. Further, $\vec p$ and $\vec p
\, '$  are the initial and final CMS momenta, $\vec \sigma_i$ ($\fet \tau_i$)
refers to the spin (isospin) matrices of the nucleon $i$ and  
$\vec q \equiv \vec p \, ' - \vec p$,
$\vec k \equiv \frac{1}{2} (\vec p \, ' + \vec p \, )$. 
The leading contributions to the 2NF due to intermediate $\Delta$ excitations
arise at NLO, $\nu = 2$, from diagrams shown in Fig.~\ref{nlo}. 
In the context of chiral EFT, they were first discussed by Ord\'o\~nez et 
al.~\citep{Ordonez:1995rz} using old-fashioned time-ordered 
perturbation theory. These contributions were then re-considered by Kaiser 
et al.~\citep{Kaiser:1998wa} using the Feynman graph technique. In that work,
compact analytical expressions for the corresponding non-polynomial pieces 
of the TPEP were presented. For the sake of completeness, we list below the 
results from \citep{Kaiser:1998wa} which are generalized to spectral-function
regularization with an arbitrary cutoff $\tilde \Lambda $ (for a precise
definition of spectral function regularization, see \cite{Epelbaum:2003gr}):
\begin{itemize}
\item
$\Delta$--excitation in the triangle graphs:
\beq
\label{del_tri}
W_C = -\frac{h_A^2}{216 \pi^2 F_\pi^4}\,
\biggl\{ (6\Sigma -\omega^2) L^{\tilde \Lambda}(q) +
12 \Delta^2 \Sigma D^{\tilde \Lambda} (q) \biggr\}\,.
\eeq
\item
Single $\Delta$--excitation in the box graphs:
\beqa
\label{del_sing}
V_{C} &=& -\frac{g_A^2 \, h_A^2}{12 \pi
  F_\pi^4 \Delta }\, (2M_\pi^2 +q^2)^2 \,  A^{\tilde \Lambda} (q) \,,\nn
W_C &=& - \frac{g_A^2 \, h_A^2}{216 \pi^2 F_\pi^4}\, 
\biggl\{(12 \Delta^2-20 M_\pi^2-11q^2) L^{\tilde \Lambda}(q) + 6\Sigma^2
  D^{\tilde \Lambda}(q) \biggr\} \,,
\nn
V_T &=&  - \frac{1}{q^2} V_S = -\frac{g_A^2 \, h_A^2}{48 \pi^2 F_\pi^4}\,  
\biggl\{ -2L^{\tilde \Lambda}(q) + (\omega^2-4 \Delta^2) D^{\tilde \Lambda}(q) \biggr\}\,, \nn
W_T &=& - \frac{1}{q^2} W_S = -\frac{g_A^2 \, h_A^2}{144 \pi F_\pi^4 \Delta}\,
  \omega^2 \, A^{\tilde \Lambda}(q)~.
\eeqa
\item
Double $\Delta$--excitation in the box graphs:
\beqa
\label{del_doub}
V_{C} &=& -\frac{h_A^4}{27 \pi^2
  F_\pi^4}\, \biggl\{-4 \Delta^2 L^{\tilde \Lambda}(q) + \Sigma [ H^{\tilde
  \Lambda}(q) + (\Sigma +8 \Delta^2)D^{\tilde \Lambda}(q) ] \biggr\}\,, \nn
W_C &=& - \frac{h_A^4}{486 \pi^2 F_\pi^4}\, 
\biggl\{(12 \Sigma - \omega^2) L^{\tilde \Lambda}(q) + 3\Sigma [H^{\tilde
  \Lambda}(q)+ (8 \Delta^2 - \Sigma )D^{\tilde \Lambda}(q)] \biggr\} \,,
\nn
V_T &=&  - \frac{1}{q^2} V_S = - \frac{h_A^4}{216 \pi^2 F_\pi^4}\,  
\biggl\{ 6L^{\tilde \Lambda}(q) + (12 \Delta^2- \omega^2) D^{\tilde \Lambda}(q) \biggr\}\,, \nn
W_T &=& - \frac{1}{q^2} W_S = - \frac{h_A^4}{1296  \pi^2 F_\pi^4} \,
  \biggl\{
2L^{\tilde \Lambda}(q) + (4 \Delta^2 + \omega^2 ) D^{\tilde \Lambda}(q) \biggr\}\,.
\eeqa
\end{itemize}
The quantities $\Sigma$, $L^{\tilde \Lambda}$, $A^{\tilde \Lambda}$,
$D^{\tilde \Lambda}$ and  $H^{\tilde \Lambda}$ in the above expressions are
defined as follows: 
\beqa
\label{definitions}
\Sigma &=&  2M_\pi^2 + q^2 -2 \Delta^2\,, \nn
L^{\tilde \Lambda} (q) &=& \theta (\tilde \Lambda - 2 M_\pi ) \, \frac{\omega}{2 q} \, 
\ln \frac{\tilde \Lambda^2 \omega^2 + q^2 s^2 + 2 \tilde \Lambda q 
\omega s}{4 M_\pi^2 ( \tilde \Lambda^2 + q^2)}~, \quad
\omega = \sqrt{ q^2 + 4 M_\pi^2}~,  \quad 
s = \sqrt{\tilde \Lambda^2 - 4 M_\pi^2}\,, \nn
A^{\tilde \Lambda} (q) &=& \theta ( \tilde \Lambda - 2 M_\pi ) \frac{1}{2 q} 
\arctan \frac{q (\tilde \Lambda - 2 M_\pi )}{q^2 + 2 \tilde \Lambda M_\pi}\,, \nn
D^{\tilde \Lambda}(q) &=& \frac{1}{\Delta} \, \int_{2M_\pi}^{\tilde \Lambda}
\frac{d\mu}{\mu^2+q^2} \, \arctan \frac{\sqrt{\mu^2-4M_\pi^2}}{2\Delta}\,, \nn
H^{\tilde \Lambda}(q) &=& \frac{2 \Sigma }{\omega^2-4 \Delta^2} \biggl[
  L^{\tilde \Lambda}(q) - L^{\tilde \Lambda} (2\sqrt{\Delta^2 -
  M_\pi^2}) \biggr] \,.
\eeqa

\begin{figure}[tb]
  \begin{center} 
\includegraphics[width=13.0cm,keepaspectratio,angle=0,clip]{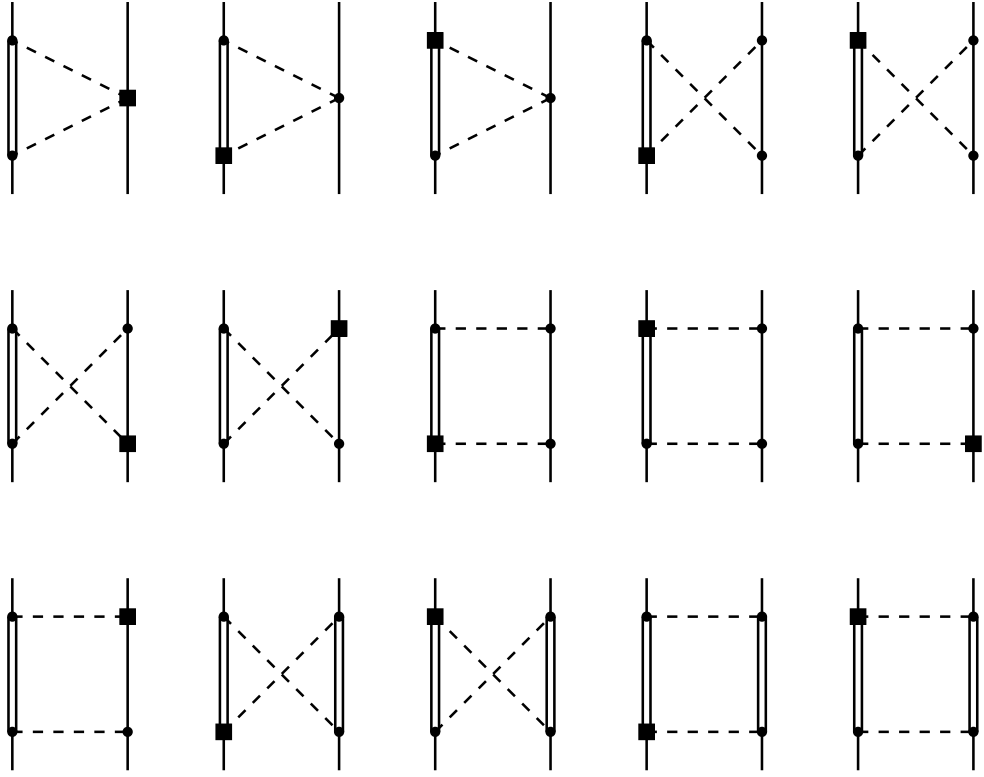}
    \caption{Subleading contributions to the $2\pi$--exchange NN potential 
with single and 
double $\Delta$--excitations. Filled squares denote  subleading vertices. 
For remaining notation see Fig.~\ref{nlo}.  
\label{nnlo} 
 }
  \end{center}
\end{figure}
The subleading contributions to the 2NF due to intermediate $\Delta$--excitations
arise at NNLO, $\nu = 3$, from the diagrams shown in Fig.~\ref{nnlo}. 
The contribution from the first graph was already discussed in 
Ref.~\cite{Ordonez:1995rz} based on
time-ordered perturbation theory. Notice that the loop integral was treated
numerically, and no contributions involving the subleading $\pi N \Delta$
vertices were considered in that work.  

The simplest way to evaluate the
subleading $\Delta$-contributions to the NN potential is using the Feynman
graph technique. Since the corresponding diagrams do not involve reducible
topologies, one obtains the same result for the corresponding static TPEP
using Feynman diagrams, time-ordered perturbation theory or the method of
unitary transformations. We find the following expressions for the subleading 
$\Delta$-contributions:
\begin{itemize}
\item
$\Delta$--excitation in the triangle graphs:
\beqa
\label{del_tri3}
V_{C} &=& -\frac{h_A^2 \, \Delta }{18 \pi^2 
  F_\pi^4}\, \biggl\{ 6 \Sigma \, \Big(4 c_1 M_\pi^2 - 2 c_2 \Delta^2 - c_3 ( 2
  \Delta^2 + \Sigma ) \Big) \, D^{\tilde \Lambda}(q) \nn
      && \mbox{\hskip 1.9 true cm} + \Big(- 24 \, c_1 M_\pi^2 + c_2 \, 
  (\omega^2 - 6 \Sigma ) + 6 \, c_3 \, ( 2 \Delta^2 + \Sigma ) \Big) L^{\tilde
  \Lambda}(q) \biggr\}  \,, \nn
W_{C} &=& -\frac{(b_3 + b_8 ) \, h_A \, \Delta }{108 \pi^2 
  F_\pi^4}\, \biggl\{ 12 \Delta^2 \, \Sigma  \, D^{\tilde \Lambda}(q) + (- \omega^2
  + 6 \Sigma )  L^{\tilde \Lambda}(q) \biggr\} \,, \nn
W_T &=& - \frac{1}{q^2} W_S = -\frac{c_4 \, h_A^2 \, \Delta }{72 \pi^2 F_\pi^4}\,
  \biggl\{ ( \omega^2 - 4 \Delta^2 )  D^{\tilde \Lambda}(q) - 2 L^{\tilde
  \Lambda}(q) \biggr\} \,. 
\eeqa
\item
Single $\Delta$--excitation in the box graphs:
\beqa
\label{del_sing3}
W_{C} &=& -\frac{(b_3 + b_8 ) \, g_A^2 \, h_A \, \Delta }{108 \pi^2 
  F_\pi^4}\, \biggl\{ 6 \Sigma^2 \,  D^{\tilde \Lambda}(q) + \Big( \omega^2 -
12 (\Delta^2 + \Sigma ) \Big)  L^{\tilde \Lambda}(q)  \biggr\} \,, \nn
V_T &=& - \frac{1}{q^2} V_S = -\frac{(b_3 + b_8 ) \, g_A^2 \, h_A \, \Delta }{24 \pi^2 
  F_\pi^4}\, \biggl\{ ( \omega^2 - 4 \Delta^2 ) D^{\tilde \Lambda}(q)  - 2
L^{\tilde \Lambda}(q) \biggr\} \,.
\eeqa
\item
Double $\Delta$--excitation in the box graphs:
\beqa
\label{del_doub3}
V_{C} &=& 6 W_C = -\frac{4 (b_3 + b_8 ) \, h_A^3 \, \Delta }{81 \pi^2 
  F_\pi^4}\, \biggl\{ 3 ( 8 \Delta^2 - \Sigma ) \, \Sigma \, 
  D^{\tilde \Lambda}(q) + 3 \Sigma \, H^{\tilde \Lambda}(q)  + 
  ( - \omega^2 + 12 \Sigma )  L^{\tilde \Lambda}(q)  \biggr\} \,, \nn
V_{T} &=& - \frac{1}{q^2} V_S = 6 W_{T} = - \frac{6}{q^2} W_S = -\frac{(b_3 +
  b_8 ) \, h_A^3 \, \Delta }{54 \pi^2 
  F_\pi^4}\, \biggl\{  (- \omega^2 + 12 \Delta^2 ) D^{\tilde \Lambda}(q)  + 6
L^{\tilde \Lambda}(q) \biggr\} \,.
\eeqa
\end{itemize}

It is instructive  to verify the consistency between the results 
obtained in chiral EFT with and without explicit $\Delta$'s. 
Clearly,  both formulations differ from each other only by the different
counting of the $\Delta N$ mass splitting, $\Delta \sim M_\pi \ll \Lambda_\chi$
versus $M_\pi \ll \Delta \sim \Lambda_\chi$. Expanding the various terms in
Eqs.~(\ref{del_tri})-(\ref{del_doub3}) in powers of $1/\Delta $  and counting $\Delta
\sim \Lambda_\chi$ should, therefore, yield either terms polynomial in momenta
(i.e.~contact interactions) or non-polynomial contributions
absorbable 
into a  redefinition of the LECs in chiral EFT without explicit
$\Delta$'s (in harmony with the decoupling theorem).
Expanding Eqs.~(\ref{del_tri})-(\ref{del_doub3}) in powers of $1/\Delta $ and
using the relations 
\beqa
D^{\tilde \Lambda}(q) &=&  - \frac{1}{2 \Delta^2} \, L^{\tilde \Lambda}(q) - 
\frac{1}{24 \Delta^4} \, (4 M_\pi^2 + q^2) \, L^{\tilde \Lambda}(q) +
\mbox{\it polyn} + \mathcal{O} \left( \Delta^{-6} \right)\,, \nn
H^{\tilde \Lambda}(q) &=&  L^{\tilde \Lambda}(q) - \frac{q^2}{4 \Delta^2} \,
L^{\tilde \Lambda}(q) +
\mbox{\it polyn} + \mathcal{O} \left( \Delta^{-4} \right)\,,
\eeqa
where {\it polyn} refers to terms  polynomial in $q^2$,
it is easy to verify that the only remaining non-polynomial contributions up to NNLO 
are the ones given in
the first and the last lines in Eq.~(\ref{del_sing}). They can be exactly reproduced 
in EFT without explicit $\Delta$'s by an appropriate shift in the LECs $c_3$
and $c_4$ \cite{Kaiser:1998wa}. The contributions to these LECs due to the
$\Delta$, $c_3 = - 2 c_4 = -{4} h_A^2/(9 \Delta )$,
were analyzed already  in \cite{Bernard:1996gq}.

\section{Determination of the LECs from pion-nucleon scattering}
\def\theequation{\arabic{section}.\arabic{equation}}
\label{sec:LEC}

\begin{figure}[tb]
  \begin{center} 
\includegraphics[width=10.0cm,keepaspectratio,angle=0,clip]{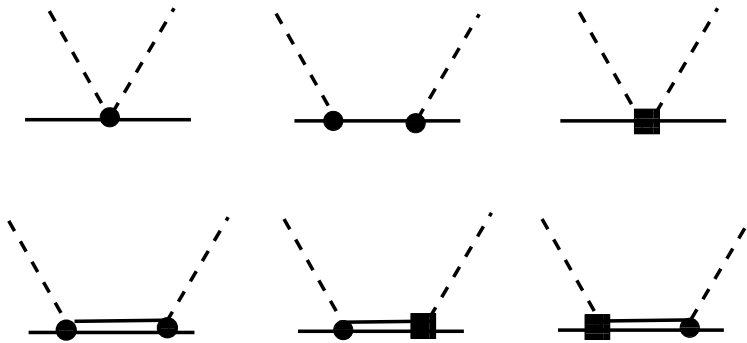}
    \caption{Tree graphs for pion-nucleon scattering at NLO. The filled
             circles/squares denote leading/subleading vertices.
             Crossed graphs are not shown. Note that due to the Weinberg
             counting of the nucleon mass, some graphs with fixed 
             dimension-two LECs do not appear at this order. Also, diagrams
             that simply lead to a nucleon mass shift are not shown.
\label{fig:diag} 
 }
  \end{center}
\end{figure}

Next, we must fix parameters. At leading order, we use $g_A = 1.27$
for the nucleon axial-vector coupling. The corresponding $N \Delta$
axial-coupling $h_A$ is less well known, we therefore use here two extreme
values, namely $h_A = 1.05$ as in Ref.~\cite{Fettes:2000bb} and $h_A =
3g_A/(2\sqrt{2}) = 1.34$ from SU(4) (or large $N_c$).
To proceed, we must determine the various dimension-two LECs.
We determine these LECs from a fit at order $Q^2$ to the S-- and P--wave 
$\pi N$ threshold parameters, which is consistent at the order we are
working. Note that because of our treatment of the nucleon mass, there are no 
$1/m$-corrections to the $\pi N$ amplitude at this order. The pertinent Feynman
diagrams describing elastic $\pi N$ scattering at this order are shown in 
Fig.~\ref{fig:diag}. 
The explicit analytical expressions for the threshold parameters
calculated at ${\cal O}(Q^2)$ are collected in App.~\ref{app:threshold}.
These depend on the $\pi N$ LECs $c_i$ ($i = 1, \ldots, 4$) and the combination
of $\pi N \Delta$ LECs  $b_3 + b_8$.  To determine these couplings, we fit to 
the EM98 phase shifts, corresponding to Fit~2 in Ref.~\cite{Fettes:2000bb}.
The resulting values for the LECs are collected Tab.~\ref{tab:ci}
and the corresponding threshold parameters are shown in Tab.~\ref{tab:a's}. 
Various remarks are
in order. First, we note that the inclusion of the $\Delta$ leads to much
smaller values for $c_2$ and $c_3$ and a sizeably reduced $c_4$. This is
consistent with the findings in \cite{Bernard:1996gq}, where it was shown
that about half of the strength of $c_4$ is due to the $\rho$-meson exchange
in the t-channel whereas $c_2$ and $c_3$ are largely given by the delta. 
The two fits with the delta, denoted fit~1 and fit~2 in
Tab.~\ref{tab:ci} differ by  the value of $h_A$ used as input. The
resulting  values of the LECs are quite different, but (as we will show later)
the TPEP does not change much.     
The combination $b_3+b_8$ comes out of natural size in both fits.
Also notice that both fits lead to very similar results for $\pi N$ threshold 
coefficients (the differences are not visible in Table \ref{tab:a's}).
We also stress that there are significant deviations between
different PWAs leading to further uncertainties in the LECs. Furthermore,
additional significant changes in the LECs appear if one goes to ${\cal O}(Q^3)$.
For these reasons, we do not assign any uncertainties to these LECs in our
study which can only be addressed with sufficient precision in a  one-loop
calculation within the SSE. What is already obvious from the values for 
the dimension-two LECs is that the TPEP contribution will be greatly enhanced/reduced at
NLO/NNLO leading to a better convergence than for the delta-less theory.
This is also reflected in the results displayed in  Tab.~\ref{tab:a's}.  
At second order in the chiral expansion one is not able to fit some of the 
threshold parameters, most  notably the P-wave scattering volumes. This 
changes drastically when the delta is included -- this is consistent with the 
statement made earlier that the inclusion of the $\Delta$ resums some
important contributions to the $\pi N$ scattering amplitude (but not all). 

\begin{table*}[t] 
\begin{center}
\begin{tabular}{||ccccc||ccccc|ccccc|ccccc||ccccc||}
\hline \hline 
&&           &&&&&          &&&&&           &&&&&         &&&&&  &&  \\[-1.5ex]
&& LECs &&&&& $Q^2$, no  $\Delta$ &&&&&  $Q^2$,  fit 1  &&&&&
$Q^2$,  fit 2 &&&&& $Q^3$ no $\Delta$ \cite{Fettes:1998ud}, fit 2   &&  \\[1ex]
\hline \hline 
&&           &&&&&          &&&&&           &&&&&         &&&&&  &&  \\[-1.5ex]
&& $c_1$     &&&&& $-$0.57  &&&&& $-$0.57   &&&&& $-$0.57 &&&&& $-1.42\pm0.03$ &&    \\[0.3ex]
&& $c_2$     &&&&& ~~2.84   &&&&& $-$0.25   &&&&& ~~0.83  &&&&& ~~$3.13 \pm 0.04$ &&  \\[0.3ex]
&& $c_3$     &&&&& $-$3.87  &&&&& $-$0.79   &&&&& $-$1.87 &&&&& $- 5.58 \pm 0.01$ &&  \\[0.3ex]
&& $c_4$     &&&&& ~~2.89   &&&&& ~~1.33    &&&&& ~~1.87  &&&&& ~~$3.50\pm 0.01$ &&   \\[0.3ex]
&& $h_A$     &&&&& --       &&&&& ~~~$1.34^\star$ &&&&& ~~~$1.05^\star$ &&&&& -- &&  \\[0.3ex]
&& $b_3+b_8$ &&&&& --       &&&&& ~~1.40    &&&&& ~~2.95    &&&&& -- &&     \\[1ex]
\hline \hline
  \end{tabular}
\caption{Determinations of the LECs from S-- and P--wave threshold parameters in $\pi N$
   scattering based on the $Q^2$ fits with and without explicit
   $\Delta$'s. LECs used as input are marked by the star.  Also shown are the
   values determined in Ref.~\cite{Fettes:1998ud} from fit 2
   at $Q^3$ without explicit $\Delta$'s (the errors given are purely
   statistical and do not reflect the true uncertainty of the LECs). 
    The LECs $c_i$ and $b_3 + b_8$ are given in GeV$^{-1}$.   
\label{tab:ci}}
\end{center}
\end{table*}

\begin{table*}[htb] 
\begin{center}
\begin{tabular}{||ccccc||ccccc|ccccc||ccccc||ccccc||}
\hline \hline 
   &&  &&&&&  &&&&& &&&&& &&&&&  &&  \\[-1.5ex]
   &&  &&&&& $Q^2$, no $\Delta$ &&&&&  $Q^2$  fits 1, 2 &&&&&
   $Q^3$ no $\Delta$ \cite{Fettes:1998ud}, fit 2  &&&&&  EM98  && \\[1ex]
\hline \hline 
   &&  &&&&&  &&&&&  &&&&&  &&&&& &&  \\[-1.5ex]
&& $a_{0 +}^+$  &&&&& ~~0.41   &&&&& ~~0.41   &&&&& ~~0.49 &&&&& $0.41 \pm 0.09$ &&  \\[0.3ex]
&& $b_{0 +}^+$  &&&&& $-$4.46  &&&&& $-$4.46  &&&&& $-$5.23 &&&&& $- 4.46$  &&  \\[0.3ex]
&& $a_{0 +}^-$  &&&&& ~~7.74   &&&&& ~~7.74   &&&&& ~~7.72 &&&&& $7.73 \pm 0.06$  && \\[0.3ex]
&& $b_{0 +}^-$  &&&&& ~~3.34   &&&&& ~~3.34   &&&&& ~~1.62 &&&&& $1.56$ &&  \\[0.3ex]
&& $a_{1 -}^-$  &&&&& $-$0.05  &&&&& $-$1.32  &&&&& $-$1.19  &&&&& $- 1.19 \pm 0.08$ &&  \\[0.3ex]
&& $a_{1 -}^+$  &&&&& $-$2.81  &&&&& $-$5.30  &&&&& $-$5.38 &&&&& $- 5.46 \pm
   0.10$ &&  \\[0.3ex]
&& $a_{1 +}^-$  &&&&& $-$6.22  &&&&& $-$8.45  &&&&& $-$8.16 &&&&& $- 8.22 \pm 0.07$ &&  \\[0.3ex]
&& $a_{1 +}^+$  &&&&& ~~9.68   &&&&& ~~12.92  &&&&& ~~13.66 &&&&& $13.13 \pm 0.13$  && \\[1ex]
\hline \hline
  \end{tabular}
\caption{Values of the S-- and P--wave threshold parameters in $\pi N$
   scattering for the various fits as described in the text on comparison with
   the data. Both fit 1 and fit 2 lead to very similar results. Units are
   appropriate  inverse powers of the pion mass times $10^{-2}$.
\label{tab:a's}}
\end{center}
\end{table*}

\section{Results for the potential and the peripheral waves}
\def\theequation{\arabic{section}.\arabic{equation}}
\label{sec:results}

Having determined the LECs we are now in the
position to present the results for the potential and peripheral partial
waves. To be precise, we first consider the various components of the NN
potential in coordinate space and then discuss the resulting peripheral
waves.

The coordinate space representations of the various components of the
TPEP up to NNLO are defined according to   
\beqa
\tilde V (r) &=& \tilde V_C (r) +  \tilde W_C (r) \; (\fet \tau_1 \cdot \fet \tau_2 ) 
+ \Bigl[  \tilde V_S (r)   +  \tilde W_S (r) \; (\fet \tau_1 \cdot \fet \tau_2 ) \Bigr]
\, (\vec \sigma_1 \cdot \vec \sigma_2) \nonumber\\
&&{} + \Bigl[  \tilde V_T (r)   +  \tilde W_T (r) \; (\fet \tau_1 \cdot \fet \tau_2 ) \Bigr]
(3 \vec \sigma_1 \cdot \hat r \; \vec \sigma_2 \cdot \hat r  - 
\vec \sigma_1 \cdot \vec \sigma_2 )\,.
\eeqa
The functions $\tilde V_{C,S,T} (r)$ can be determined 
for any given $r > 0$ via
\beqa
\label{fourc}
\tilde V_{C} (r) &=&  \frac{1}{2 \pi^2 r} \int_{2 M_\pi}^\infty d \mu \, \mu \, 
e^{-\mu r} \,\rho_{C} (\mu)\,, \nonumber \nn
\label{fourt}
\tilde V_{T} (r) &=& -\frac{1}{6 \pi^2 r^3}  \int_{2 M_\pi}^\infty d \mu \, \mu \, 
e^{-\mu r} \, ( 3 + 3 \mu r + \mu^2 r^2 ) \rho_T (\mu)\,, \nonumber \nn
\label{fours}
\tilde V_{S} (r) &=&  -\frac{1}{6 \pi^2 r}  \int_{2 M_\pi}^\infty d \mu \, \mu \, 
e^{-\mu r} \, \Bigl( \mu^2 \rho_T (\mu) - 3 \rho_S (\mu ) \Bigr)\,,  
\eeqa
where the spectral functions $\rho_i (\mu )$ are obtained from $V_i (q )$ by
\cite{Kaiser:1997mw}:
\beq
\rho_i (\mu ) = {\rm Im}~\left[ V_i (0^+ - i \mu ) \right]\,.
\eeq
The coordinate space representations of the isovector parts are given 
by the above equations replacing 
$\tilde V_{C,S,T} (r) \to \tilde W_{C,S,T} (r)$ and 
$V_{C,S,T} (q) \to  W_{C,S,T} (q)$. Notice further that for 
$\mu > 2 M_\pi$:
\beqa
{\rm Im}~\left[ A^{\tilde \Lambda} (0^+ - i \mu ) \right] &=& \frac{\pi}{ 4 \mu } \, 
\theta (\tilde \Lambda - \mu ) \,, \nn 
{\rm Im}~\left[ L^{\tilde \Lambda} (0^+ - i \mu ) \right] &=& - \frac{ \pi }{ 2 \mu }
\, \sqrt{\mu^2 - 4 M_\pi^2} \, 
\theta (\tilde \Lambda - \mu )\, , \nn 
{\rm Im}~\left[ D^{\tilde \Lambda} (0^+ - i \mu ) \right]  &=& \frac{\pi}{2 \mu \Delta} \arctan
\frac{\sqrt{\mu^2 - 4 M_\pi^2}}{2 \Delta} \, \theta (\tilde \Lambda - \mu )\,.
\eeqa

\begin{figure}[tb]
  \begin{center} 
\includegraphics[width=16.6cm,keepaspectratio,angle=0,clip]{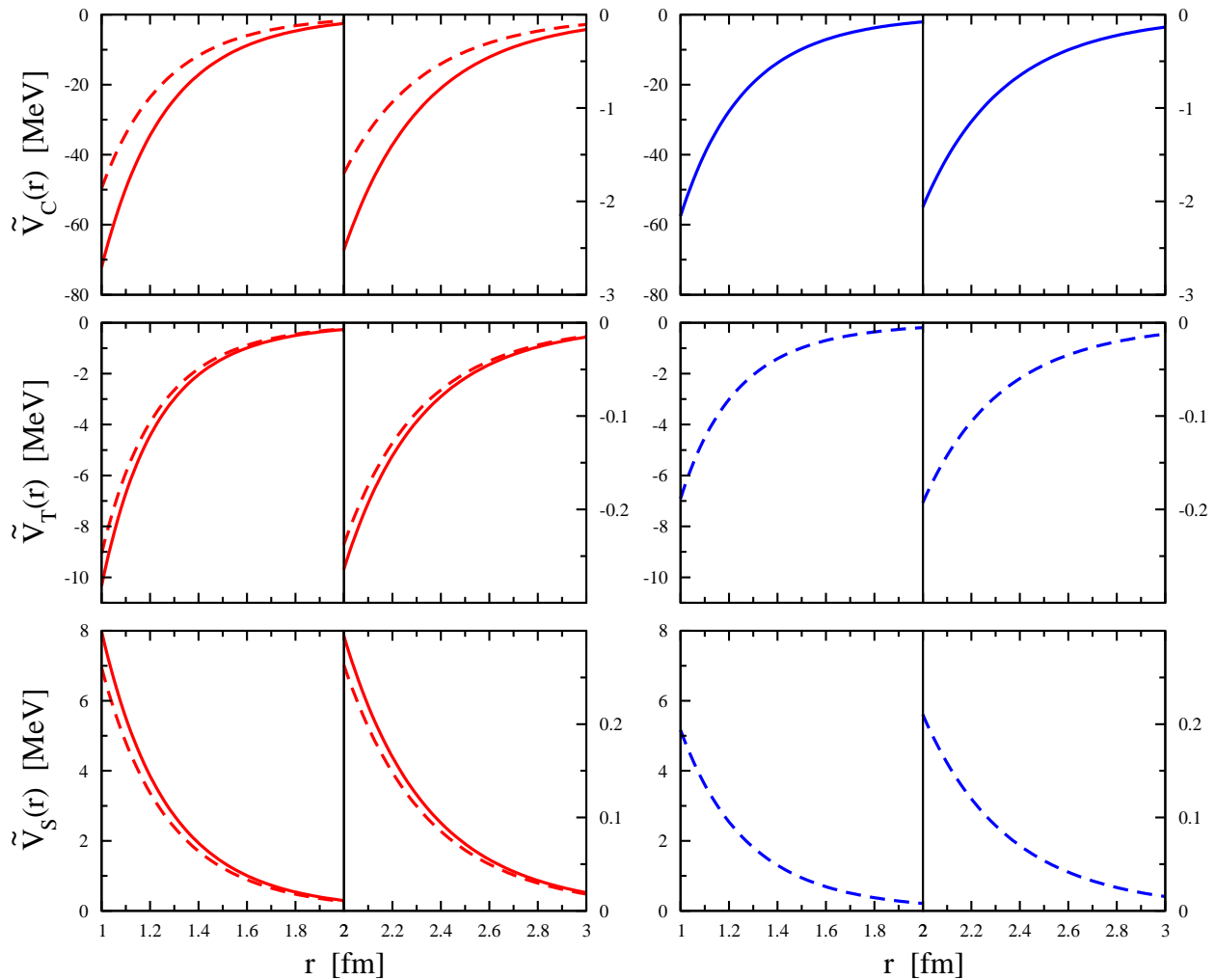}
    \caption{Isoscalar two-pion exchange potentials in coordinate space. Left
      (right) panels show the results with (without) explicit
      $\Delta$'s. Dashed and solid lines refer to the NLO and NNLO results,
      respectively. There are no contributions to $\tilde V_C$ ($\tilde V_{T,S}$) at
      NLO-$\Delta \palka$ (NNLO-$\Delta \palka$).
\label{v} 
 }
  \end{center}
\end{figure}

\begin{figure}[htb]
  \begin{center} 
\includegraphics[width=16.6cm,keepaspectratio,angle=0,clip]{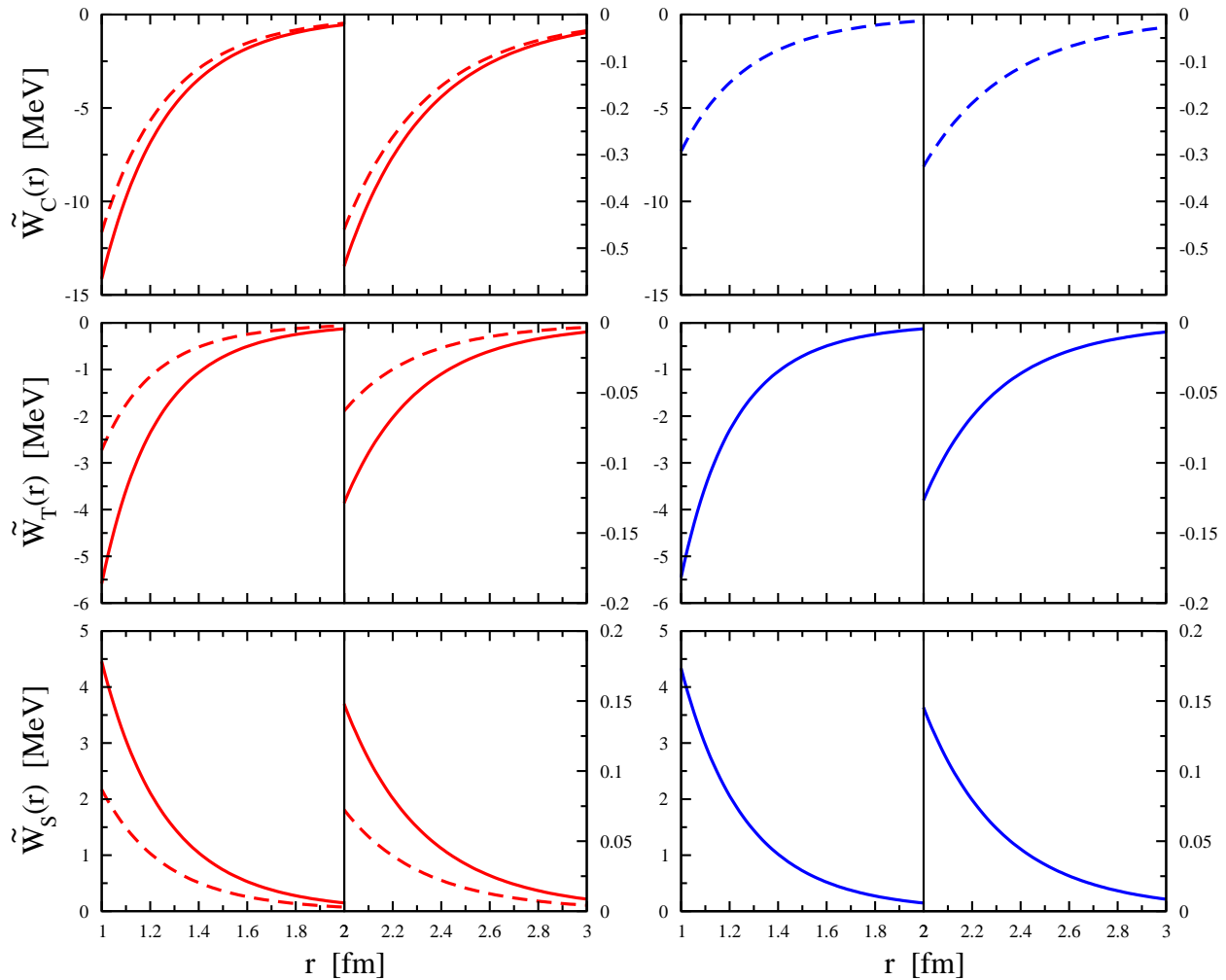}
    \caption{Isovector two-pion exchange potentials in coordinate space. Left
      (right) panels show the results with (without) explicit
      $\Delta$'s. Dashed and solid lines refer to the NLO and NNLO results,
      respectively. There are no contributions to $\tilde W_C$ ($\tilde W_{T,S}$) at
      NNLO-$\Delta \palka$ (NLO-$\Delta \palka$).
\label{w} 
 }
  \end{center}
\end{figure}

\begin{figure}[htb]
  \begin{center} 
\includegraphics[width=16.6cm,keepaspectratio,angle=0,clip]{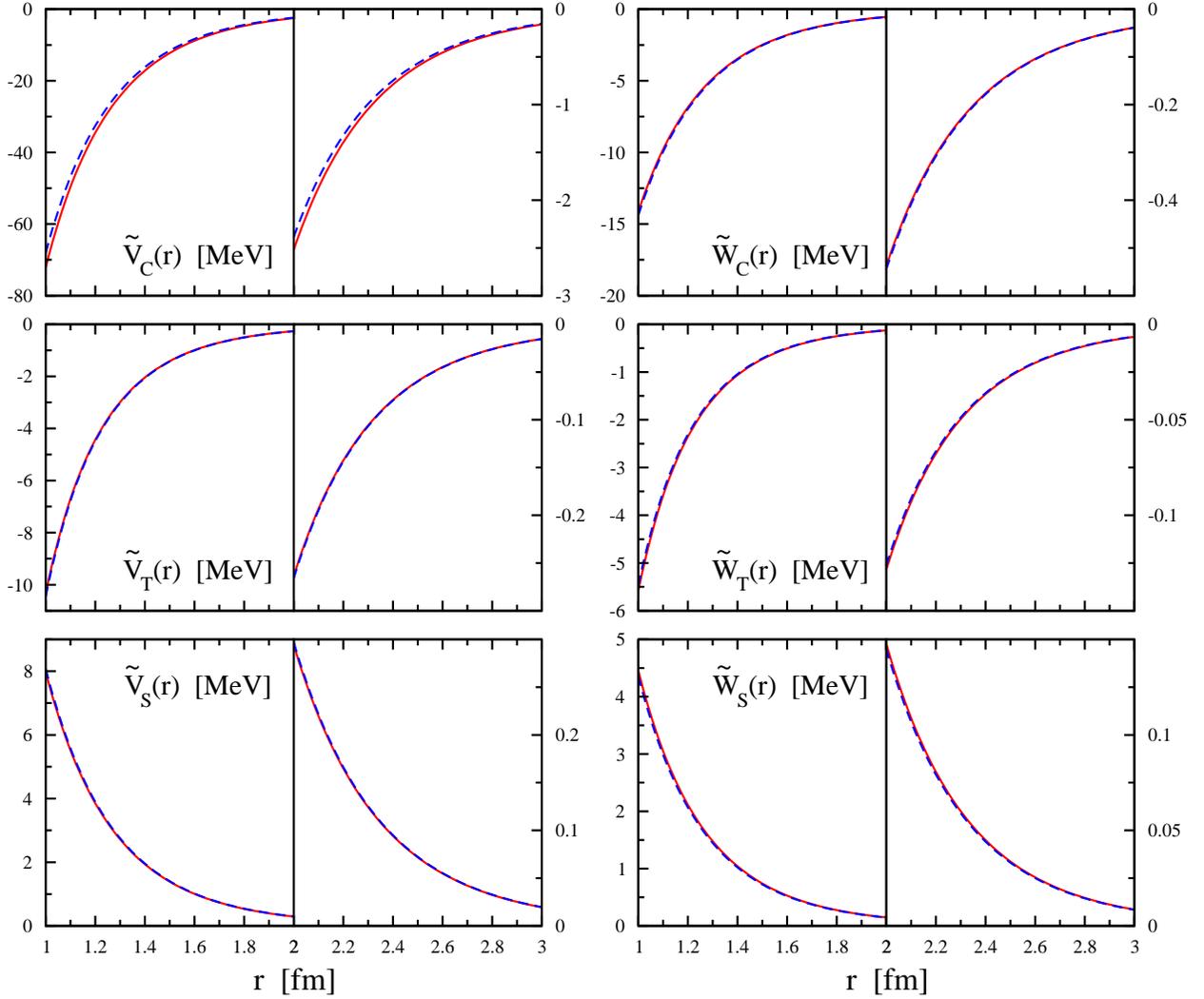}
    \caption{Sensitivity of the NNLO two-pion exchange potential to the choice
      of the low-energy constant $h_A$. Solid (dashed) lines show the NNLO
      potentials with $h_A = {3} g_A/(2 \sqrt{2}) \sim 1.34$ ($h_A = 1.05$) 
      and the LECs $c_i$ chosen according to fits 1 and 2 in Tab.~\ref{tab:ci}.
\label{vw_delta} 
 }
  \end{center}
\end{figure}

In Figs.~\ref{v} and \ref{w} the results for the functions $\tilde V_{C,S,T}
(r)$ and $\tilde W_{C,S,T} (r)$ entering the TPEP are shown up to NNLO in EFT with
(EFT-$\Delta$) and without explicit $\Delta$'s (EFT-$\Delta \palka$) and using
 $\tilde \Lambda = 700$ MeV.  
For the LECs $c_i$ and $b_3 + b_8$, we use the values obtained in fit 1, see
Tab.~\ref{tab:ci}.    
The results at NNLO-$\Delta \palka$ are based on the values for $c_i$ given  
in the second column of Tab.~\ref{tab:ci} (these are not very different from
the ones obtained in the literature using other values of these LECs). 
The isoscalar central potential
$\tilde V_C (r)$ is known to be the strongest part of the chiral 
TPEP. In EFT-$\Delta \palka$, it receives the first contribution at NNLO. 
The SSE leads to a much more natural convergence pattern for $\tilde V_C (r)$ 
with the dominant contribution being generated at NLO-$\Delta$. The correction at 
NNLO-$\Delta$ is still significant and produces further attraction. Notice
also that the NNLO-$\Delta \palka$ result for $\tilde V_C (r)$  lies between
the NLO-$\Delta$ and NNLO-$\Delta$ ones. Contrary to the central part, the
isoscalar tensor and spin-spin potentials $\tilde V_{T,S} (r)$ as well as the
isovector central potential $\tilde W_{C} (r)$ receive the contributions at 
NLO-$\Delta \palka$ with no further corrections at NNLO-$\Delta \palka$. 
As shown in Figs.~\ref{v} and \ref{w}, the leading contributions 
due to the intermediate $\Delta$-excitation turn out to be significant in all
these cases while the subleading corrections (i.e. at  NNLO-$\Delta$) are less 
important. Notice that the resulting potential $\tilde W_{C} (r)$ at NNLO-$\Delta$ is
nearly twice as strong as at  NNLO-$\Delta \palka$. The strength of the remaining
isovector tensor and spin-spin parts of the potential $\tilde W_{T,S} (r)$
at NNLO-$\Delta \palka$ comes out remarkably close to the one obtained at the
same order in the EFT without explicit $\Delta$'s, see Fig.~\ref{w}. Again,
contrary to EFT-$\Delta \palka$, the NLO contributions to $\tilde W_{T,S} (r)$
in the theory with explicit $\Delta$'s do not vanish. 
Let us now comment on the sensitivity of the TPEP to the value of the LEC $h_A$
used as an input in the determination of the LECs $c_i$ and 
$b_3 + b_8$, see the discussion in section \ref{sec:LEC}. While the values
of the LECs obtained in fits 1 and 2 for two different choices of $h_A$ differ
significantly from each other, see Tab.~\ref{tab:ci}, the resulting S- and
P-wave threshold parameters nearly coincide in both cases, see Tab.~\ref{tab:a's}.       
As demonstrated in Fig.~\ref{vw_delta}, a similar trend holds also for the
TPEP. 
The only visible (but still very small) difference is observed for the
isoscalar central part of the potential. 

We will now consider D- and higher partial waves up to NNLO in chiral EFT
with and without explicit deltas following the lines of 
Refs.~\citep{Kaiser:1998wa,Epelbaum:2003gr}. 
Using Born approximation for the scattering amplitude, 
the phase shifts and mixing angles in the convention of Stapp et
al.~\citep{Stapp:1956mz} 
are determined by the 2N potential $V$ as:
\beq
\delta^{sj}_l = - \frac{m q}{(4 \pi)^2}  \langle lsj | V | lsj \rangle \,, \quad \quad
\epsilon_j = - \frac{m q}{(4 \pi)^2}  \langle j-1,sj | V | j+1,sj \rangle \,,
\eeq
where $s$ and $j$ refer to the total spin and angular momentum, respectively,
and $q$ is the nucleon CMS momentum. 
Clearly, such an approximation violates unitarity. 
This violation is small provided that the corresponding phase shifts are small. Alternatively,
one can use the $K$--matrix approach in order to restore unitarity, see e.g.~\citep{Entem:2002sf}. 
Here and in what follows, we adopt the same notation for the matrix elements
in the $|lsj \rangle$ basis as 
in Ref.~\citep{Epelbaum:2004fk}. The expressions for the partial wave
decomposition can be found in appendix B of this reference. 

\begin{figure}[htb]
  \begin{center} 
\includegraphics[width=14.0cm,keepaspectratio,angle=0,clip]{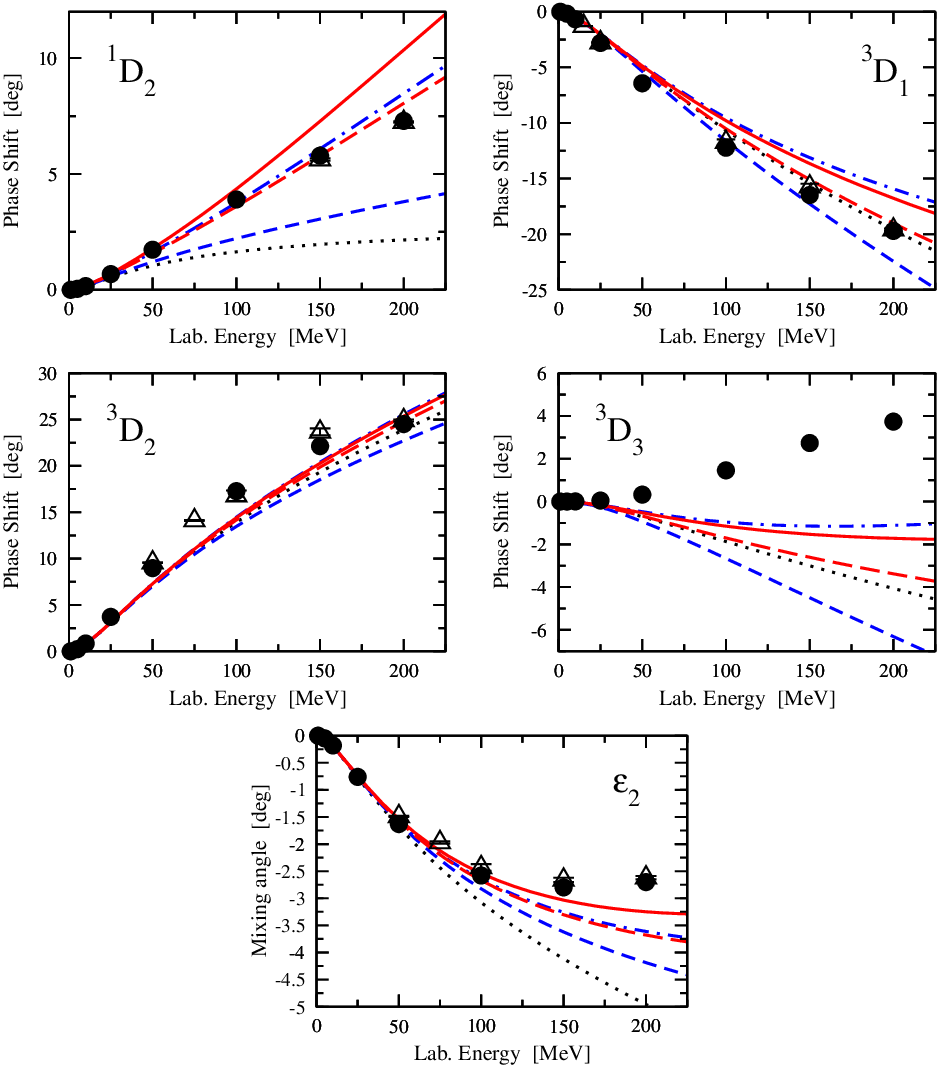}
    \caption{D--wave NN phase shifts and the mixing parameter $\epsilon_2$. 
The dotted curve is the LO prediction (i.e.~based on the pure OPEP).
Long-dashed (short-dashed) and solid (dashed-dotted) lines show the NLO and
NNLO results with (without) the explicit $\Delta$-contributions and using the
SFR with $\tilde \Lambda =700$ MeV. The filled circles (open triangles) depict
the results from the Nijmegen multi--energy PWA \citep{Stoks:1993tb,NNonline} 
(Virginia Tech single--energy PWA \citep{SAID}). 
\label{dwaves} 
 }
  \end{center}
\end{figure}

\begin{figure}[htb]
  \begin{center} 
\includegraphics[width=14.0cm,keepaspectratio,angle=0,clip]{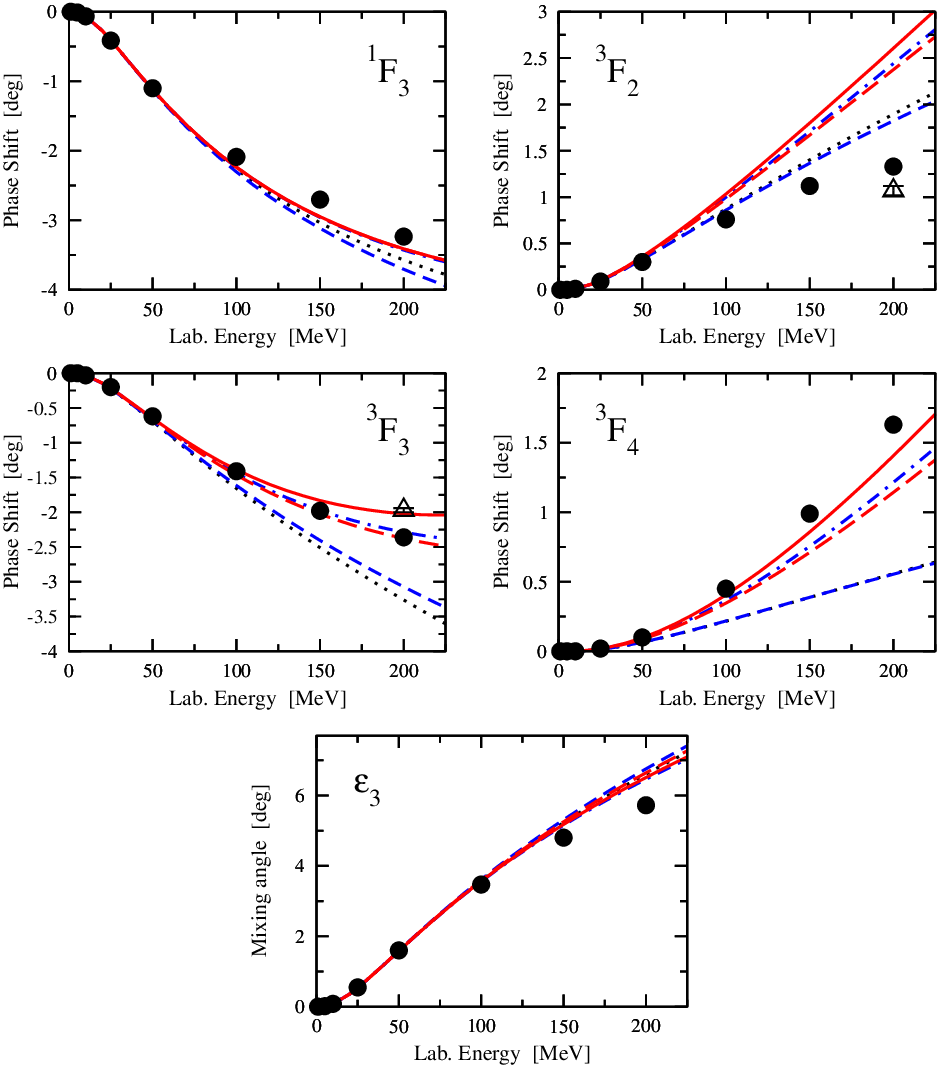}
    \caption{F--wave NN phase shifts and the mixing parameter $\epsilon_3$. 
The dotted curve is the LO prediction (i.e.~based on the pure OPEP).
Long-dashed (short-dashed) and solid (dashed-dotted) lines show the NLO and
NNLO results with (without) the explicit $\Delta$-contributions and using the
SFR with $\tilde \Lambda =700$ MeV. The filled circles (open triangles) depict
the results from the Nijmegen multi--energy PWA \citep{Stoks:1993tb,NNonline} 
(Virginia Tech single--energy PWA \citep{SAID}). 
\label{fwaves} 
 }
  \end{center}
\end{figure}

\begin{figure}[htb]
  \begin{center} 
\includegraphics[width=14.0cm,keepaspectratio,angle=0,clip]{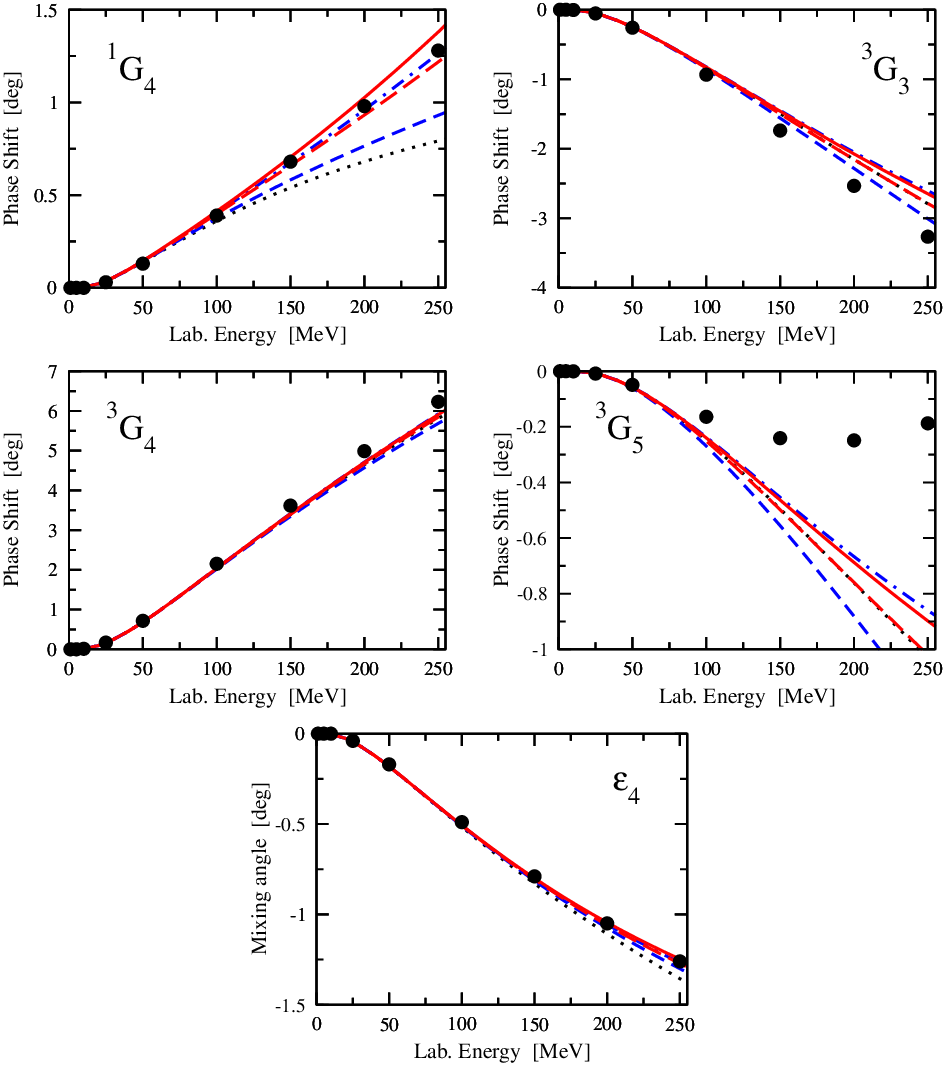}
    \caption{G--wave NN phase shifts and the mixing parameter $\epsilon_4$. 
The dotted curve is the LO prediction (i.e.~based on the pure OPEP).
Long-dashed (short-dashed) and solid (dashed-dotted) lines show the NLO and
NNLO results with (without) the explicit $\Delta$-contributions and using the
SFR with $\tilde \Lambda =700$ MeV. The filled circles depict
the results from the Nijmegen multi--energy PWA \citep{Stoks:1993tb,NNonline}. 
\label{gwaves} 
 }
  \end{center}
\end{figure}

Our results for D- F- and G-waves and for the mixing parameters
$\epsilon_{2,3,4}$ are shown in Figs.~\ref{dwaves}-\ref{gwaves}.
The two sets of the LECs obtained in fit 1 and fit 2, see Tab.~\ref{tab:ci},
yield nearly the same results for all phase shifts and mixing parameters as one
might expect from Fig.~\ref{vw_delta}. 
In all channels, the convergence pattern is strongly improved in the SSE as compared to
the pure chiral expansion. While the NLO corrections are small and the
dominant contributions in the EFT-$\Delta \palka$ arise at NNLO, the changes 
between NLO and NNLO are typically much smaller than the ones between LO and NLO in the 
EFT-$\Delta$.  The only exceptions are observed in the $^3D_1$, $^3D_3$ and
$^3G_5$, where the NLO-$\Delta$ result appears to be close to the LO one. 
In all other partial waves, the
results at NLO-$\Delta$ are very similar to the ones at NNLO-$\Delta \palka$. 
While most peripheral partial waves are reasonably well described at both 
NNLO-$\Delta$ and  NNLO-$\Delta \palka$ using the SFR cutoff $\tilde \Lambda = 700$
MeV, the description of the $^3D_3$ and $^3G_5$ phase shifts is rather poor. 
It should, however, be understood that since the phase shifts in these
channels are much smaller in magnitude than in the other D- and G-waves,
respectively, the observed deviations have little effect on NN scattering
observables. Notice also that, especially in the
$^3D_3$ partial wave, the contributions 
due to iteration of the potential in the Lippmann-Schwinger equation,
which are not considered in this work, might be significant.   

\begin{figure}[htb]
  \begin{center} 
\includegraphics[width=14.0cm,keepaspectratio,angle=0,clip]{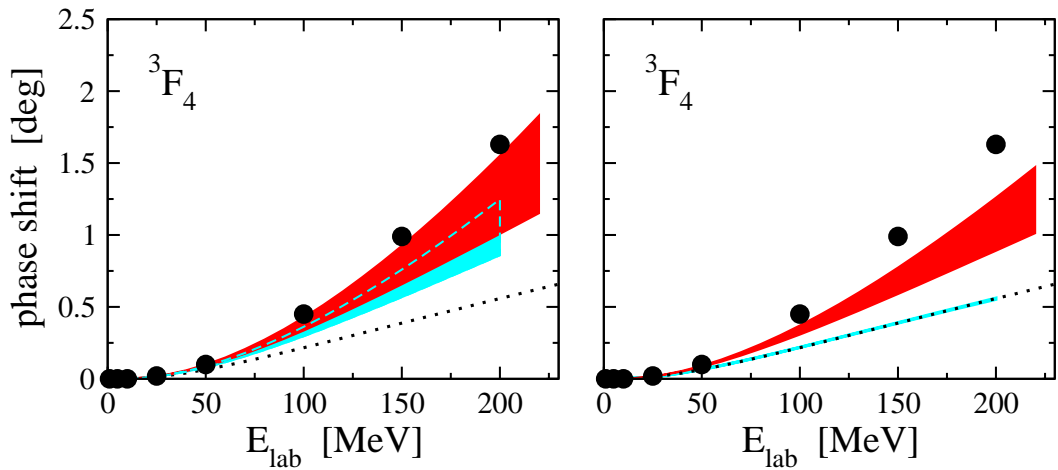}
    \caption{Theoretical uncertainty in the $^3F_4$ partial wave due to the
      variation of the SFR cutoff as described in the text. Left panel:
    Delta-full theory. The upper (lower) band refers to NNLO (NLO) and the
    dotted line is the leading order OPEP. Right panel: Same for the delta-less theory.
\label{3f4} 
 }
  \end{center}
\end{figure}

Finally, we would like to emphasize that the results shown in
Figs.~\ref{dwaves}-\ref{gwaves} correspond to one particular choice of the SFR
cutoff $\tilde \Lambda = 700$ MeV. The theoretical uncertainty associated with the 
variation of $\tilde \Lambda$ in the range $\tilde \Lambda = 500 \ldots 800$ MeV is
exemplified in Fig.\ref{3f4} where the $^3F_4$ partial wave is shown at
NNLO-$\Delta$ and NNLO-$\Delta \palka$. We note that the uncertainty at NNLO
is comparable in both approaches. Note further that the results for
NNLO-$\Delta \palka$ differs slightly from the one displayed in Ref.~\cite{Epelbaum:2003gr}
due to a different choice of the LECs $c_{3,4}$.
For a detailed discussion on the theoretical
uncertainty in both peripheral and low partial waves in EFT without explicit
$\Delta$'s the reader
is referred to \cite{Epelbaum:2003gr,Epelbaum:2003xx,Epelbaum:2004fk}.

\section{Summary and conclusions}
\def\theequation{\arabic{section}.\arabic{equation}}
\label{sec:summary}

In this paper, we have analyzed the two-nucleon two-pion-exchange
potential in an effective field theory with explicit deltas (based on 
the so-called small scale expansion) at next-to-next-to-leading order 
in the chiral expansion. The pertinent results can be summarized as follows:
\begin{itemize}
\item[i)]At NNLO, the potential with single and double delta-excitations
is shown in Figs.~\ref{nlo},\ref{nnlo}. It depends on 7 low-energy constants,
the leading order axial-vector couplings $g_A$ and $h_A$, the dimension-two
$\pi N$ LECs $c_1, c_2, c_3, c_4$ and the combination $b_3+b_8$ from the
$\pi N \Delta$ Lagrangian.
\item[ii)]We have determined these dimension-two LECs from a fit to 
the pion-nucleon S-and P-wave threshold parameters, based on 
the ${\cal O}(Q^2)$ representation of the $\pi N$ scattering amplitude
in the SSE, see Tables~\ref{tab:ci},\ref{tab:a's}. As expected, the values
for the $c_i$ ($i=2,3,4$) are considerably smaller than in  chiral
perturbation theory.
\item[iii)]The resulting coordinate space potentials show a more natural
convergence behaviour than in the theory without deltas. In the delta-full
theory, the leading contributions to the TPEP at NLO turn out to be significant 
while the subleading NNLO corrections are smaller, see Figs.~\ref{v},\ref{w}.
\item[iv)]This convergence pattern is also reflected in the peripheral partial
waves, that have been calculated in first Born approximation,
see Figs.~\ref{dwaves}--\ref{gwaves}. Furthermore, for
most partial waves the description at NLO in the delta-full theory is better
than in the delta-less theory. We also find a mild improvement in some partial
waves at NNLO. For the larger D-waves, these results should
only be considered indicative due to the accuracy of the Born approximation. 
\item[v)]The theoretical uncertainty due to the variation of the SFR cutoff
$\tilde\Lambda$ is of the expected size and comparable to the case of the delta-less
theory at NNLO, see Fig.~\ref{3f4} for one typical case.
\end{itemize} 

The work presented here paves the way to a systematic analysis of nuclear
forces based on a theory with explicit deltas. A lot of work remains to be
done before this will be achieved, in particular:
\begin{itemize}
\item[-] The low partial waves and the deuteron require a non-perturbative 
treatment, i.e.~the solution of the regularized Lippmann-Schwinger  equation. 
This will be dealt with in a subsequent paper \cite{kem2}.
\item[-] To achieve a higher precision, we have to go to N$^3$LO. To do
that, one must first reconsider $\pi N$ scattering within the SSE at
one-loop order, extending the work of Ref.~\cite{Fettes:2000bb}. It also 
requires a careful re-evaluation of the corresponding two- and 
three-pion exchange contributions with
intermediate deltas. That these corrections will not be negligible follows
from the calculations of sub-leading three-pion-exchanges in the delta-less
theory, see Ref.~\cite{Kaiser:2001dm}.
\item[-] In the theory with explicit deltas, three-nucleon force (3NF) arises
already at NLO. Its leading term is included in all current 3NF models and is
usually referred to as the  Fujita-Miyazawa force~\cite{FM50}. 
The first corrections to it appear at NNLO, some of these are again 
proportional to the combination of LECs $b_3 + b_8$. The ordering of various
contributions to the 3NF in the theory with explicit deltas 
is therefore very different from the one based on the pure pion-nucleon
theory, as discussed e.g. in~\cite{Pandharipande:2005sx}.
\item[-] Additional isospin violation effects arise in the 2NF and 3NF due
to the mass splittings of the various delta charge states. These should be
investigated systematically within the consistent EFT extending earlier 
model-dependent work~\cite{Li:1998xh}.
\end{itemize}
\section*{Acknowledgments}

The work of E.E. and H.K. was supported in parts by funds provided from the 
Helmholtz Association to the young investigator group  
``Few-Nucleon Systems in Chiral Effective Field Theory'' (grant  VH-NG-222). 
This work was further supported by the DFG (SFB/TR 16 ``Subnuclear Structure
of Matter'') and by the EU Integrated Infrastructure Initiative Hadron
Physics Project under contract number RII3-CT-2004-506078.

\appendix

\def\theequation{\Alph{section}.\arabic{equation}}
\setcounter{equation}{0}
\section{{\boldmath$\pi N$} threshold coefficients at order {\boldmath$Q^2$}}
\label{app:threshold}

In this appendix we give explicit expressions for $\pi N$ threshold
coefficients $a_{0+}^{\pm}$, $b_{0+}^{\pm}$ and $a_{1\pm}^{\pm}$  at order
$Q^2$ (calculated in the SSE) used in section \ref{sec:LEC} 
to determine the corresponding LECs.  These are in standard notation:
\beqa
a_{0+}^+ &=& \frac{1}{2 \pi F_\pi^2 ( m + M_\pi )
  }\, ( -2 c_1 + c_2 + c_3 ) \, m M_\pi^2 \,, \nn
b_{0+}^+ &=& \frac{1}{8 \pi F_\pi^2 m ( m + M_\pi )
  }\, \Big( 2 c_1 ( 2 m - M_\pi ) M_\pi + c_2 (4 m^2 - 2 m M_\pi + M_\pi^2
  ) + c_3 (4 m^2 - 2 m M_\pi + M_\pi^2 ) \Big) \,, \nn
a_{0+}^- &=& \frac{1}{8 \pi F_\pi^2 ( m + M_\pi )
  }\, m M_\pi \,, \nn
b_{0+}^- &=& \frac{1}{32 \pi F_\pi^2 m M_\pi ( m + M_\pi )
  }\, (2 m^2 - 2 m M_\pi + M_\pi^2 ) \,, \nn
a_{1-}^- &=& -\frac{1}{24 \pi F_\pi^2 M_\pi ( m + M_\pi )
  }\, m \, (g_A^2 - 4 c_4 M_\pi ) + \frac{1}{27 \pi F_\pi^2 ( m + M_\pi )
  (M_\pi + \Delta ) } \, m \, \Big(h_A^2  - 2 h_A \, (b_3 + b_8 ) M_\pi \Big)\,, \nn
a_{1-}^+ &=& -\frac{1}{12 \pi F_\pi^2 M_\pi ( m + M_\pi )
  }\, m \, (g_A^2 + 2 c_3 M_\pi ) + \frac{2}{27 \pi F_\pi^2 ( m + M_\pi )
  (M_\pi + \Delta ) } \, m \, \Big(h_A^2  - 2 h_A \, (b_3 + b_8 ) M_\pi
\Big)\,, \nn
a_{1+}^- &=& -\frac{1}{24 \pi F_\pi^2 M_\pi ( m + M_\pi )
  }\, m \, (g_A^2 + 2 c_4 M_\pi ) \nn
&& {}+ \frac{1}{54 \pi F_\pi^2 ( m + M_\pi )
  (M_\pi^2 - \Delta^2 ) } \, m \, \Big(h_A^2 (2 M_\pi + \Delta ) + 2 h_A \,
(b_3 + b_8 ) M_\pi (M_\pi + 2 \Delta )
\Big)\,, \nn
a_{1+}^+ &=& \frac{1}{24 \pi F_\pi^2 M_\pi ( m + M_\pi )
  }\, m \, (g_A^2 - 4 c_3 M_\pi ) \nn
&& {}- \frac{1}{27 \pi F_\pi^2 ( m + M_\pi )
  (M_\pi^2 - \Delta^2 ) } \, m \, \Big(h_A^2 (M_\pi + 2 \Delta ) + 2 h_A \,
(b_3 + b_8 ) M_\pi (2 M_\pi + \Delta )
\Big)\,.
 \eeqa

\setlength{\bibsep}{0.2em}

\end{document}